\documentclass[12pt]{article}%
\usepackage{amsmath}
\usepackage{graphicx}
\usepackage{amsfonts}
\usepackage{amssymb}%
\providecommand{\U}[1]{\protect\rule{.1in}{.1in}}
\begin{document}

\title{Numerical Calculation of the Fidelity for the Kondo and the Friedel-Anderson Impurities}
\author{Gerd Bergmann and Richard S. Thompson\\Department of Physics \& Astronomy\\University of Southern California\\Los Angeles, California 90089-0484\\e-mail: bergmann@usc.edu}
\date{\today}
\maketitle

\begin{abstract}
The fidelities of the Kondo and the Friedel-Anderson (FA) impurities are
calculated numerically. The ground states of both systems are calculated with
the FAIR (Friedel artificially inserted resonance) theory. The ground state in
the interacting systems is compared with a nullstate in which the interaction
is zero. The different multi-electron states are expressed in terms of Wilson
states. The use of $N$ Wilson states simulates the use of a large effective
number $N_{eff}$ of states. A plot of $\ln(F)$ versus $N\varpropto\ln\left(
N_{eff}\right)  $ reveals whether one has an Anderson orthogonality
catastrophe at zero energy. The results are at first glance surprising. The
$\ln\left(  F\right)  -\ln\left(  N_{eff}\right)  $ plot for the Kondo
impurity diverges for large $N_{eff}$. On the other hand, the corresponding
plot for the symmetric FA impurity saturates for large $N_{eff}$ when the
level spacing at the Fermi level is of the order of the singlet-triplet
excitation energy. The behavior of the fidelity allows one to determine the
phase shift of the electron states in this regime.

PACS: 75.20.Hr, 71.23.An, 71.27.+a , 05.30.-d

\newpage

\end{abstract}

\section{Introduction}

In the process of modeling a complicated physical state by a simplified model
it is of great interest how well the model agrees with the real state. To
measure this agreement one compares the two states with each other. If the
states are electronic wave functions then the comparison can be performed as a
scalar product between the two wave functions. The result is called the
fidelity, and it is defined as%
\begin{equation}
F=|\left\langle \Psi_{\text{model}}|\Psi_{\text{real}}\right\rangle |
\end{equation}
It turns out that this concept is also useful when the system (for example the
Hamiltonian) depends on a parameter $\lambda$. Then one can define the
fidelity as
\[
F=\left\vert \left\langle \Psi_{\lambda}|\Psi_{0}\right\rangle \right\vert
\]
This definition is slightly different from the definition of the differential
fidelity $F\left(  \lambda,d\lambda\right)  $
\begin{equation}
F\left(  \lambda,d\lambda\right)  =\left\vert \left\langle \Psi_{\lambda}%
|\Psi_{\lambda+d\lambda}\right\rangle \right\vert =1-\frac{1}{2}G\left(
\delta\lambda\right)  ^{2}%
\end{equation}
where $G$ is the fidelity susceptibility \cite{Z32}, \cite{Z31}, \cite{B200},
\cite{Z33}, \cite{S80}.

If for example a potential in the Hamiltonian is given by $\lambda V,$ then
when the potential $\lambda V$ acts in the whole volume (as for example in the
periodic Hubbard model) the fidelity susceptibility is generally proportional
to the number of band electrons. (For phase transitions such as quantum
critical points it can increase faster than linearly with the number of band
electrons). Therefore it is of interest how the fidelity of a system depends
on the number of conduction electrons.

The definition of the fidelity is connected with the Anderson orthogonality
catastrophe (AOC) as introduced by Anderson \cite{A53}. Anderson showed that
the ground state of a system of $N$ fermions is orthogonal to the ground state
in the presence of a finite-range scattering potential, as $N$ approaches
infinity $\ln\left(  F\right)  \varpropto-\ln\left(  N\right)  $. This AOC has
been intensively studied in connection with the Kondo effect \cite{S91},
\cite{A94}, \cite{S90}, \cite{H32}, \cite{Y6}, \cite{Y7}, \cite{S92} where a
magnetic d-impurity interacts with the conduction electrons through an
exchange interaction $J\mathbf{s\cdot S,}$ where $\mathbf{s}$ and $\mathbf{S}$
are the spins of the conduction electrons and the d-impurity.

In this paper we study the fidelity for the Kondo and the Friedel-Anderson
(FA) impurities. Both systems are known to possess a singlet ground state. For
sufficiently large Coulomb repulsion between the spin-up and down impurity
state the FA impurity shows a behavior that is very similar to the Kondo
impurity. Schrieffer and Wolff \cite{S31} showed that in the range of a local
moment the FA Hamiltonian can be transformed into a Kondo Hamiltonian plus a
number of additional terms $\widehat{H}_{i}$. Therefore it suggestive that the
fidelities of the two systems should behave similarly. That is the reason why
we choose both systems for our investigation.

For the calculation of the fidelity we divide the Hamiltonian into two parts:
there is a part $\widehat{H}^{\lambda=0}$ that is kept constant and a second
part $\widehat{H}^{\lambda}$ that is varied during the calculation. In many
cases the Hamiltonian $\widehat{H}^{\lambda}$ depends on several parameters.
The FA impurity is an example. In this case one can choose different paths in
the parameter space. The paths are called a fidelity paths in which the
parameters $\lambda$ describe the position on the paths.

For the numerical evaluation we use the ground state which we obtain with the
FAIR (Friedel Artificially Inserted Resonance) theory \cite{B92}, \cite{B93},
\cite{B151}, \cite{B152}, \cite{B153}, \cite{B177}, \cite{B178}, \cite{B181},
\cite{B182}.

\section{Theoretical Background}

\subsection{Kondo impurity}

The Kondo system consists of a band of free s-electrons and a d-impurity with
spin $S=1/2$. Its Hamiltonian is given by%
\[
\widehat{H}^{K}=\widehat{H}^{0}+\widehat{H}^{ex}%
\]
with%
\[
\widehat{H}^{\lambda=0}=\widehat{H}^{0}=%
{\displaystyle\sum_{\nu=1}^{N}}
\varepsilon_{\nu}\widehat{c}_{\nu,\sigma}^{\dagger}\widehat{c}_{\nu,\sigma}%
\]
For the Kondo impurity we replace $\lambda$ by $J$.%
\[
\widehat{H}^{ex}=\widehat{H}^{J}=v_{a}2J\left(
{\displaystyle\sum_{\alpha,\beta}}
\widehat{\Psi}_{\alpha}^{\dag}\left(  0\right)  \mathbf{s}_{\alpha,\beta
}\widehat{\Psi}_{\beta}\left(  0\right)  \right)  \cdot\mathbf{S}%
\]%
\begin{equation}
=v_{a}J\left[
\begin{array}
[c]{c}%
\left(  S_{+}\widehat{\Psi}_{\downarrow}^{\dag}\left(  0\right)  \widehat
{\Psi}_{\uparrow}\left(  0\right)  +S_{-}\widehat{\Psi}_{\uparrow}^{\dag
}\left(  0\right)  \widehat{\Psi}_{\downarrow}\left(  0\right)  \right) \\
+S_{z}\left(  \widehat{\Psi}_{\uparrow}^{\dag}\left(  0\right)  \widehat{\Psi
}_{\uparrow}\left(  0\right)  -\widehat{\Psi}_{\downarrow}^{\dag}\left(
0\right)  \widehat{\Psi}_{\downarrow}\left(  0\right)  \right)
\end{array}
\right]  \label{Hsd}%
\end{equation}
where $S_{+},S_{-},S_{z}$ are the spin operators of the impurity with spin
$S=1/2,$ $\widehat{\Psi}_{\alpha}^{\dag}\left(  0\right)  $ and $\widehat
{\Psi}_{\beta}\left(  0\right)  $ represent field operators of the conduction
electrons and $\mathbf{s}_{\alpha,\beta}$ are the components of the Pauli
operators $\mathbf{\sigma}$ divided by two. The product $v_{a}J\widehat{\Psi
}_{\sigma}^{\dagger}\left(  0\right)  \widehat{\Psi}_{\sigma^{\prime}}\left(
0\right)  $ yields an energy since $\widehat{\Psi}_{\sigma}^{\dagger}\left(
0\right)  \widehat{\Psi}_{\sigma^{\prime}}\left(  0\right)  $ has the
dimension of a density. The operators $\widehat{c}_{\nu,\sigma}^{\dagger}$ and
$\widehat{c}_{\nu,\sigma}$ are the creation and annihilation operators for the
Wilson states of free electrons (see appendix)

The FAIR ground state for the Kondo Hamiltonian is
\begin{align}
\Psi_{K}  &  =\left[  B\widehat{a}_{0,\uparrow}^{\dag}\widehat{d}_{\downarrow
}^{\dag}+C\widehat{d}_{\uparrow}^{\dag}\widehat{b}_{0,\downarrow}^{\dag
}\right]
{\textstyle\prod\limits_{i=1}^{n-1}}
\widehat{a}_{i,\uparrow}^{\dag}%
{\textstyle\prod\limits_{j=1}^{n-1}}
\widehat{b}_{j,\downarrow}^{\dag}\Phi_{0}\label{Psi_K}\\
&  +\left[  C^{\prime}\widehat{b}_{0,\uparrow}^{\dag}\widehat{d}_{\downarrow
}^{\dag}+B^{\prime}\widehat{d}_{\uparrow}^{\dag}\widehat{a}_{0,\downarrow
}^{\dag}\right]
{\textstyle\prod\limits_{i=1}^{n-1}}
\widehat{b}_{i,\uparrow}^{\dag}%
{\textstyle\prod\limits_{j=1}^{n-1}}
\widehat{a}_{j,\downarrow}^{\dag}\Phi_{0}\nonumber
\end{align}
Here the states $\widehat{a}_{0}^{\dagger}$ and $\widehat{b}_{0}^{\dagger}$
are two artificial resonance states. The second part of the state (lower line)
is essentially the spin-reversed first part (after it is spin-ordered). In the
ground state one has $B^{\prime}=B$ and $C^{\prime}=C$ and one of the
coefficients, for example $B$, is much larger than the other so that the
relative occupations differ by a factor of about 100. Therefore the FAIR state
$\widehat{a}_{0}^{\dagger}$ is roughly the always quoted s-electron state that
forms a singlet state with the d-impurity. Details of the ground state energy
and the spatial polarization and density is discussed in \cite{B153},
\cite{B177}, \cite{B178} .

\subsection{Friedel-Anderson impurity}

For real d-electrons one has an on-site Coulomb repulsion of the d-electrons
among each other. This is described for the Friedel-Anderson (FA) impurity by
a simplified Hamiltonian%
\begin{equation}
\widehat{H}_{FA}^{\prime}=%
{\textstyle\sum_{\sigma}}
\left\{  \sum_{\nu=0}^{N-1}\varepsilon_{\nu}\widehat{c}_{\nu\sigma}^{\dag
}\widehat{c}_{\nu\sigma}+\sum_{\nu=0}^{N-1}V_{\nu}^{sd}[\widehat{d}_{\sigma
}^{\dag}\widehat{c}_{\nu\sigma}+\widehat{c}_{\nu\sigma}^{\dag}\widehat
{d}_{\sigma}]+E_{d}\widehat{d}_{\sigma}^{\dagger}\widehat{d}_{\sigma}\right\}
+Un_{d\uparrow}n_{d\downarrow} \label{H_FA}%
\end{equation}
The fact that a d-impurity has five different orbital states is simplified
into the non-degenerate case with only one d-state with spin up and another
one with spin down. The $\widehat{c}_{\nu,\sigma}^{\dagger}$ are the creation
operators for conduction electrons with spin $\sigma$ and $\widehat{d}%
_{\sigma}^{\dagger}$ is the corresponding operator for the d-electron with
spin $\sigma$. Further $V_{\nu}^{sd}$ is the matrix element for a transition
between the conduction electron $\widehat{c}_{\nu,\sigma}^{\dagger}$ and the
d-electron $\widehat{d}_{\sigma}^{\dagger}$. The most intensively studied case
is the symmetric FA impurity where $E_{d}=-U/2$.

The FAIR ground state of the FA impurity is
\begin{align}
\Psi_{SS}  &  =\left[  A\widehat{a}_{0,\uparrow}^{\dagger}\widehat
{b}_{0,\downarrow}^{\dagger}+B\widehat{a}_{0,\uparrow}^{\dagger}\widehat
{d}_{\downarrow}^{\dagger}+C\widehat{d}_{\uparrow}^{\dagger}\widehat
{b}_{0,\downarrow}^{\dagger}+D\widehat{d}_{\uparrow}^{\dagger}\widehat
{d}_{\downarrow}^{\dagger}\right]  \prod_{i=1}^{n-1}\widehat{a}_{i,\uparrow
}^{\dagger}\prod_{i=1}^{n-1}\widehat{b}_{i,\downarrow}^{\dagger}\Phi
_{0}\label{Psi_SS}\\
&  +\left[  A^{\prime}\widehat{b}_{0,\uparrow}^{\dagger}\widehat
{a}_{0,\downarrow}^{\dagger}+C^{\prime}\widehat{b}_{0,\uparrow}^{\dagger
}\widehat{d}_{\downarrow}^{\dagger}+B^{\prime}\widehat{d}_{\uparrow}^{\dagger
}\widehat{a}_{0,\downarrow}^{\dagger}+D^{\prime}\widehat{d}_{\uparrow
}^{\dagger}\widehat{d}_{\downarrow}^{\dagger}\right]  \prod_{i=1}%
^{n-1}\widehat{b}_{i,\uparrow}^{\dagger}\prod_{i=1}^{n-1}\widehat
{a}_{i,\downarrow}^{\dagger}\Phi_{0}\nonumber
\end{align}
In the ground state the coefficients $X^{\prime}=X$ where $X$ stands for
$A,B,C,D$. Again the second line is essentially the first line with reversed spins.

\section{Numerical Evaluation}

Since any solution of the Kondo or FA impurity has to include states with very
small energy (less than the Kondo energy with typical values of $10^{-5}$ or
$10^{-6}$ in units of the bandwidth) we use Wilson states (see appendix) as
the basis of our calculation. The smallest level separation at the Fermi level
for $N$ Wilson states is $\delta E=2\ast2^{-N/2}$. This energy is essential in
the fidelity calculation. A spectrum with equidistant levels would contain
$N_{eff}=2/\delta E$ states. For $N=48$ the effective number of states
$N_{eff}$ would be $N_{eff}=2/\delta E=2^{N/2}$ which is $2^{24}%
\approx1.\,\allowbreak7\times10^{7}$. This shows that with a moderate number
of Wilson states one simulates a large number of band electrons.

\subsection{Kondo impurity}

For the Kondo impurity the Hamiltonians $\widehat{H}^{\lambda=0}$ and
$\widehat{H}^{\lambda}$ have the form%

\[
\widehat{H}^{\lambda=0}=%
{\displaystyle\sum_{\nu=1}^{N}}
\varepsilon_{\nu}\widehat{c}_{\nu,\sigma}^{\dagger}\widehat{c}_{\nu,\sigma}%
\]%
\[
\widehat{H}^{\lambda}=\lambda\widehat{H}^{ex}%
\]
For comparison the state with $\lambda=0$ is required. We call this state the
nullstate. We choose for the nullstate
\[
\Psi^{\lambda=0}=\frac{1}{\sqrt{2}}\left(  \widehat{c}_{n,\uparrow}^{\dagger
}\widehat{d}_{\downarrow}^{\dagger}+\widehat{d}_{\uparrow}^{\dagger}%
\widehat{c}_{n,\downarrow}^{\dagger}\right)
{\displaystyle\prod\limits_{\nu=1}^{n-1}}
\widehat{c}_{\nu,\uparrow}^{\dagger}%
{\displaystyle\prod\limits_{\nu=1}^{n-1}}
\widehat{c}_{\nu,\downarrow}^{\dagger}\Phi_{0}%
\]
with $n=\left(  N/2+1\right)  $. This represents a half-filled band for the
spin-up and down conduction electrons plus a pseudo-singlet state between the
d-electron $\widehat{d}^{\dagger}$ and the first electron state above the
Fermi level. We call it a pseudo-singlet state because there is no coupling
between $\widehat{c}_{n}^{\dagger}$ and $\widehat{d}^{\dagger}$ since $\lambda
J$ is zero. The two components are two degenerate ground states, and their
combination represents the symmetry of the Kondo ground state.

In the next step the numerical FAIR ground states are calculated for a given
value of $\lambda J$ for a total number of Wilson states of $N=$ 20, 24, 28,
32, 36, 40, 44 and 48. Then the scalar product (the fidelity $F$) between the
nullstate and the FAIR ground state of the FA impurity is calculated. In Fig.1
the logarithm of the fidelity $\ln\left(  F\right)  $ is plotted for different
$J$ as a function of the number of Wilson states $N$. As we pointed out above
the number of Wilson states $N$ corresponds to an effective number of
electrons $N_{eff}$. With $N=2\ast\log_{2}N_{eff}$ a plot of $\ln\left(
F\right)  $ versus $N$ corresponds to log-log-plot between $F$ and $N_{eff}$.

\begin{align*}
&
{\includegraphics[
height=3.2594in,
width=3.9767in
]%
{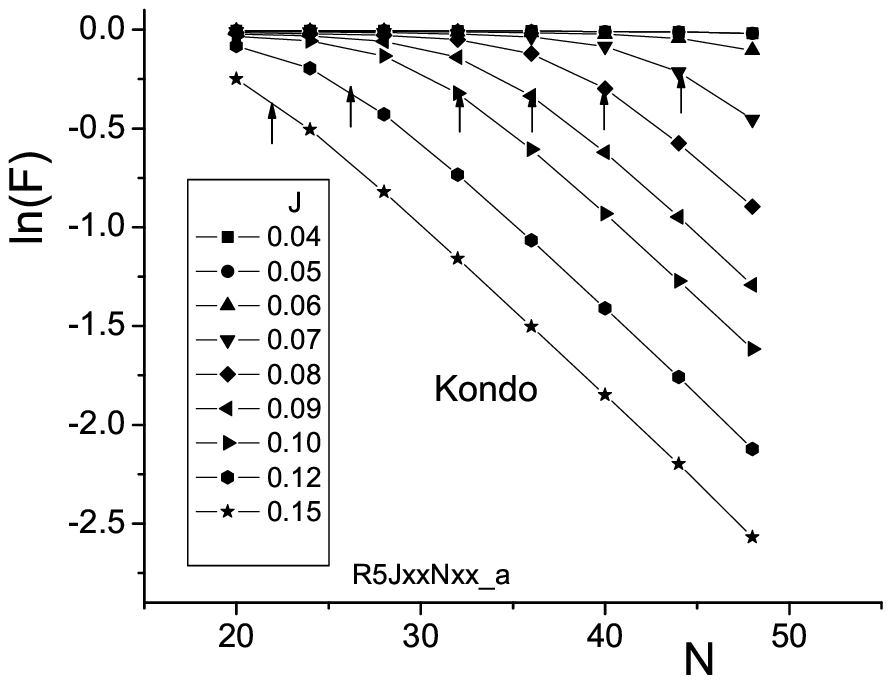}%
}%
\\
&
\begin{tabular}
[c]{l}%
Fig.1: The logarithm of the fidelity $\ln\left(  F\right)  $ for a Kondo
impurity is plotted\\
versus the number of Wilson states $N$. The nullstate for $J=0$ is described\\
in the text. (The arrows are explained in the discussion).
\end{tabular}
\end{align*}%
\[
\]

One obtains a set of curves that show in principle a linear dependence of
$\ln\left(  F\left(  0,J\right)  \right)  $ on $N$ at large values of $N$. For
the $J=0.15$ and $0.12$ curves the linear behavior is dominant for most of the
region shown. With decreasing values of $J$ the onset of the linear range
moves to larger values of $N$. For $J=0.05$ and $0.04$ the linear part is
outside of the calculated and drawn regime. In the linear regime all curves
show the same slope of $m=0.088$. So we observe that the fidelity depends on
the effective number of states as%
\[
\ln\left(  F\left(  0,J\right)  \right)  \varpropto-0.088N\approx
-0.088\ast2\log_{2}\left(  N_{eff}\right)  \approx-0.25\ln\left(
N_{eff}\right)
\]
or%
\[
F\left(  0,J\right)  \varpropto\frac{1}{N_{eff}^{1/4}}%
\]

If we consider the differential fidelity between $J=0.09$ and $0.10$ then one
obtains an interesting result that is shown in Fig.2. For small numbers of
Wilson states $F\left(  0.09,0.10\right)  $ is close to one. Then it decreases
for $N$ between $25$ and $35$ and assumes a constant value of about $0.95$ for
larger $N$. Fig.2 demonstrates very nicely that the slopes of the $\ln\left(
F\right)  $ versus $N$ curves are the same at sufficiently large $N$ (for
$J=0.09$ and $0.1$). The additional states close to the Fermi energy have the
same phase shift as we will discuss below. Fig.2 also shows that the internal
structure of the Kondo impurity for these $J$-values experiences a relative
change at $N\approx30$. Below we discuss that this is in the range of the
singlet-triplet excitation energy for the two $J$-values.%
\begin{align*}
&
{\includegraphics[
height=3.1905in,
width=3.9767in
]%
{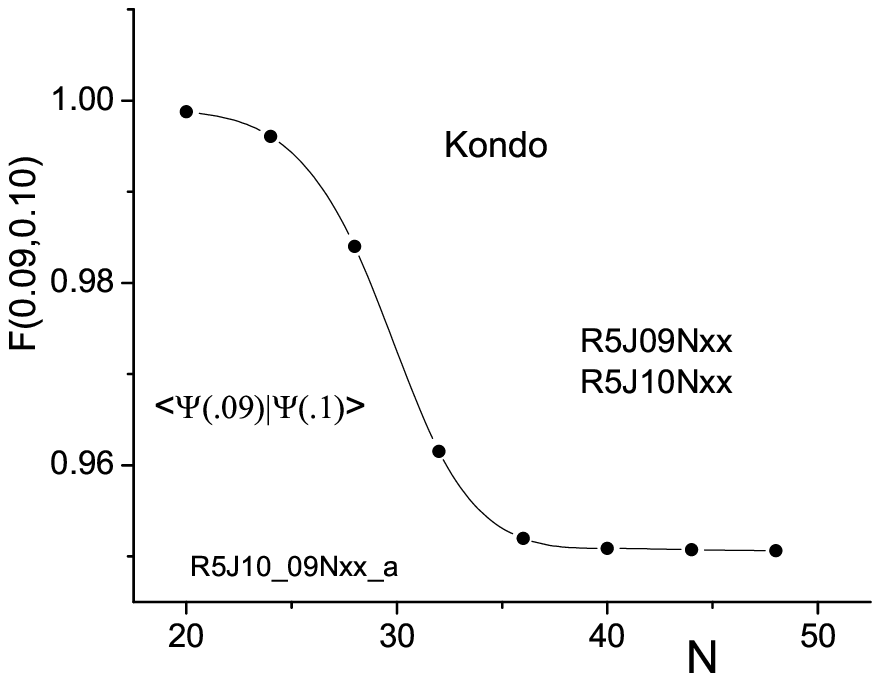}%
}%
\\
&
\begin{tabular}
[c]{l}%
Fig.2: The relative fidelity $F$ between Kondo impurities with\\
$J=0.09$ and $J=0.10$ is plotted as a function of $N$.
\end{tabular}
\end{align*}

\subsection{Friedel-Anderson impurity}

The FA impurity is described by several independent parameters, the
s-d-hopping matrix element $\left\vert V^{sd}\right\vert ^{2}$, the energy of
the d-state $E_{d}$ and the exchange energy $U$. Therefore one can choose many
different fidelity paths. In this investigation we consider essentially two
different paths; (i) the symmetric FA impurity case with $\left\vert
V^{sd}\right\vert ^{2}=0.05$, $U=\lambda U_{0}$ $\left(  U_{0}=1\right)  $ and
$E_{d}=-\frac{1}{2}\lambda U_{0},$ (ii) the asymmetric FA impurity with
constant $\left\vert V^{sd}\right\vert ^{2}=0.05$ and $U=1$ and varying
$E_{d}$ in the range $\left(  -1<E_{d}<0\right)  $.

\textbf{Symmetric case:} Here we use for the nullstate the parameters
$E_{d}=0$ and $U=0$. This represents the Friedel resonance with the d-energy
at the Fermi level. Fig.3 shows a several examples of the fidelity for
$E_{d}=-0.5$ and $U=1$ where $\left\vert V_{sd}\right\vert ^{2}$ takes the
values $0.05,0.04,0.03$ and $0.025$.%

\begin{align*}
&
{\includegraphics[
height=3.2461in,
width=3.9493in
]%
{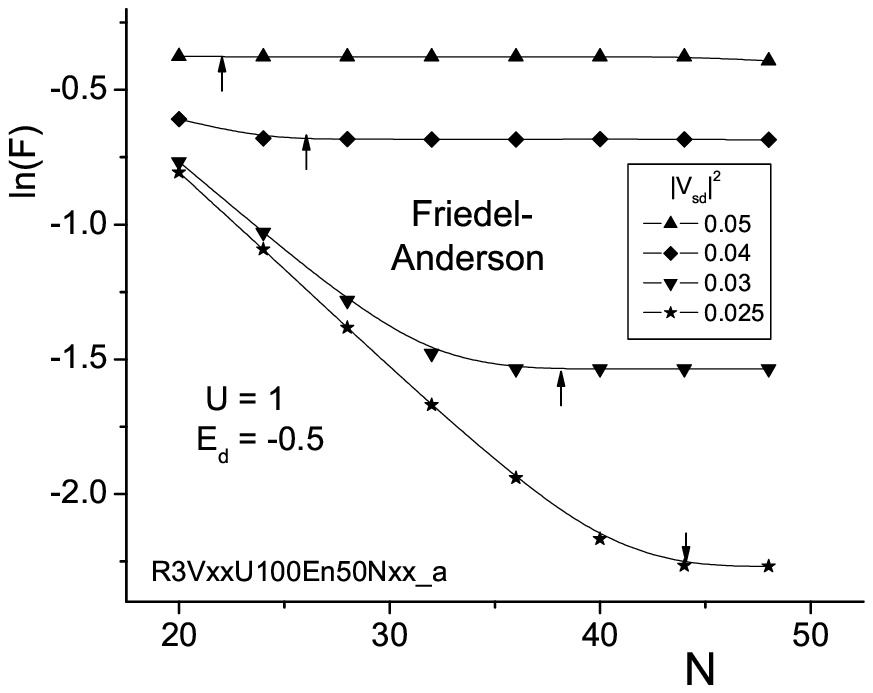}%
}%
\\
&
\begin{tabular}
[c]{l}%
Fig.3: The logarithm of the fidelity $\ln\left(  F\right)  $ for the
symmetric\\
Friedel-Anderson impurity is plotted versus the number of Wilson\\
states $N$. The nullstate for $U=0$ is the symmetric Friedel impurity\\
with the $\left\vert V_{sd}\right\vert ^{2}$ and $E_{d}=0$.
\end{tabular}
\end{align*}

Fig.3 shows that the fidelity is essentially constant for $\left\vert
V_{sd}\right\vert ^{2}=0.05$ but decreases for $\left\vert V_{sd}\right\vert
^{2}=0.025$ with increasing $N$. However, for large $N$ the fidelity
approaches a constant value. (The arrows in Fig.3 are explained in the discussion).

The values of the fidelity for $\left\vert V_{sd}\right\vert ^{2}=0.05$, $U=1$
and $E_{d}=-0.5$ vary over the whole range of $20\leq N$ $\leq48$ by less than
$2\%$. This independence of the fidelity of the number of Wilson states is
observed in the whole range $0<\lambda<1$ with $U=\lambda U_{0}$ and
$E_{d}=-\frac{\lambda}{2}U_{0}$ $\left(  U_{0}=1\right)  $. In addition the
fidelity shows a quadratic dependence on $\lambda$ as is shown in Fig.4 for
$N=32$. We observe the relationship
\[
F\left(  \lambda\right)  =1-\frac{1}{2}G\lambda^{2}%
\]
where $G$ the fidelity susceptibility has the value of $G=0.63$. There is no
unusual or singular behavior of the fidelity in the symmetric case.%
\begin{align*}
&
{\includegraphics[
height=3.1756in,
width=4.0324in
]%
{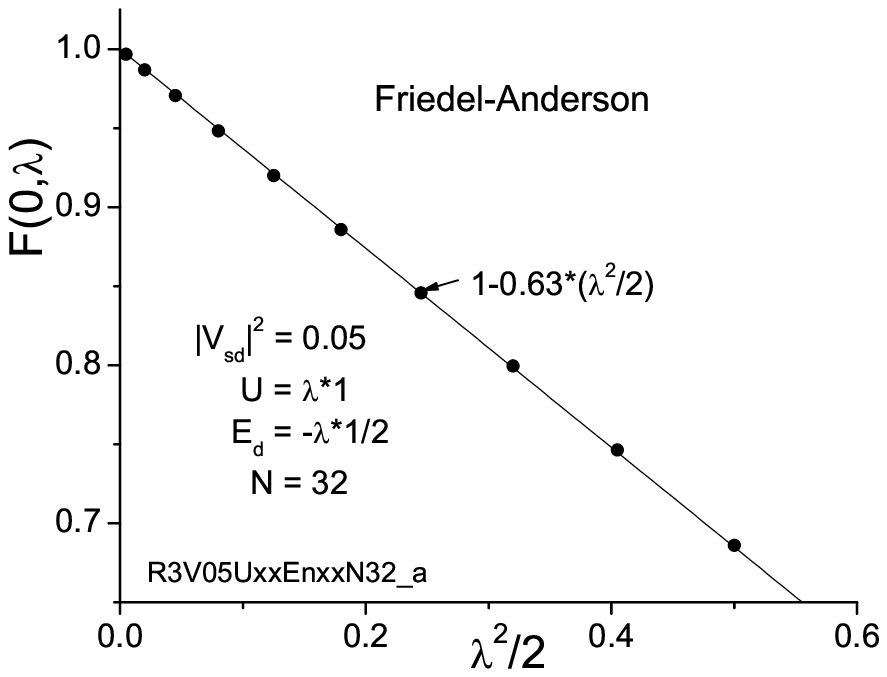}%
}%
\\
&
\begin{tabular}
[c]{l}%
Fig.4: The fidelity for $N=32$ along the path $E_{d}=-\frac{\lambda}{2}%
U_{0},U=\lambda U_{0}$\\
with $U_{0}=1$ and $\left\vert V_{sd}\right\vert ^{2}=0.05$ for a symmetric FA
impurity as a\\
function of $\lambda^{2}/2$. The nullstate is again a Friedel impurity with\\
$E_{d}=0$ and $\left\vert V_{sd}\right\vert ^{2}=0.05$.
\end{tabular}
\end{align*}

\textbf{Asymmetric case:} Next we study the fidelity of the FA impurity along
the path with $\left\vert V^{sd}\right\vert ^{2}=0.05$, $U=1$ while $E_{d}$ is
varied between $-1$ and $0$. As the nullstate we use the symmetric state with
$\left\vert V^{sd}\right\vert ^{2}=0.05$, $U=1$ and $E_{d}=-0.5$. Fig.5 shows
a typical diagram of the fidelity $\ln\left(  F\right)  $. It shows a
relatively small reduction of $\ln\left(  F\right)  $ with $N$. The small
deviation at $N=48$ is due to the fact that the calculation of the FA ground
state requires a very large number of iterations to optimize the low energy
states close to the Fermi energy. The ground-state energy has to be optimized
up to an accuracy better than $10^{-12}$. This high accuracy is normally not
needed for any other physical properties.
\begin{align*}
&
{\includegraphics[
height=3.3167in,
width=4.1029in
]%
{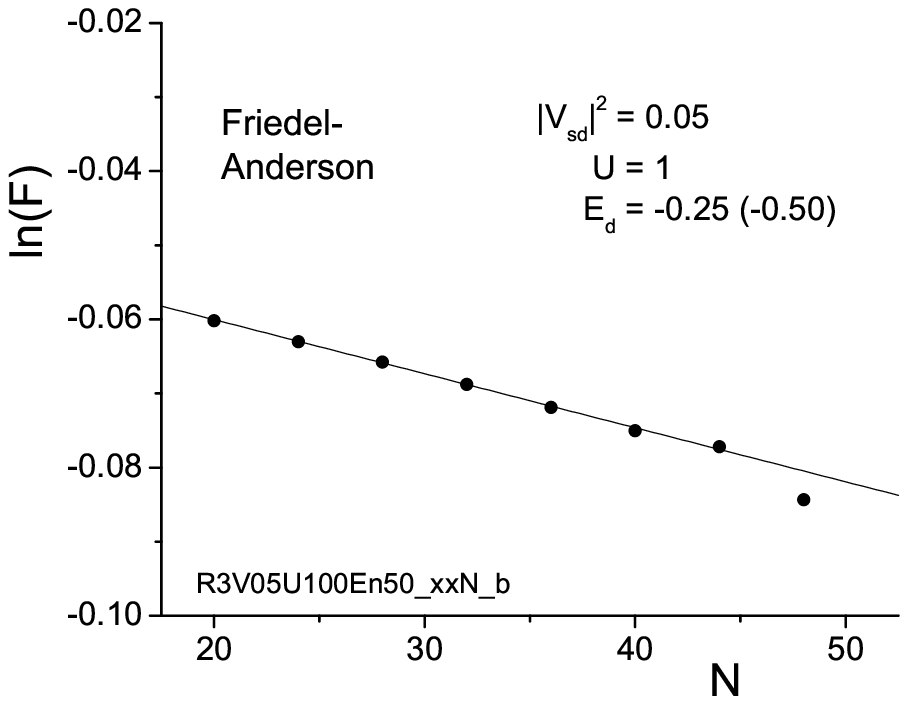}%
}%
\\
&
\begin{tabular}
[c]{l}%
Fig.5: The logarithm of the fidelity $\ln\left(  F\right)  $ between two\\
FA impurities as a function of $N$ . The impurities possess the same\\
$\left\vert V_{sd}\right\vert ^{2}=0.05$ and $U=1$ but possess different
values of $E_{d}=-0.5$\\
and $E_{d}=-0.25$.
\end{tabular}
\end{align*}

The slope $d\left(  \ln F\right)  /dN$ is shown in Fig.6 as a function of
$E_{d}$. This slope is, of course, zero at $E_{d}=-0.5$ because here the
fidelity state and the null state are identical. These results show that in
the asymmetric case one observes a reduction of the fidelity for large number
of states $N_{eff}$ but the effect is much smaller than in the Kondo
impurity.
\begin{align*}
&
{\includegraphics[
height=3.3723in,
width=4.1859in
]%
{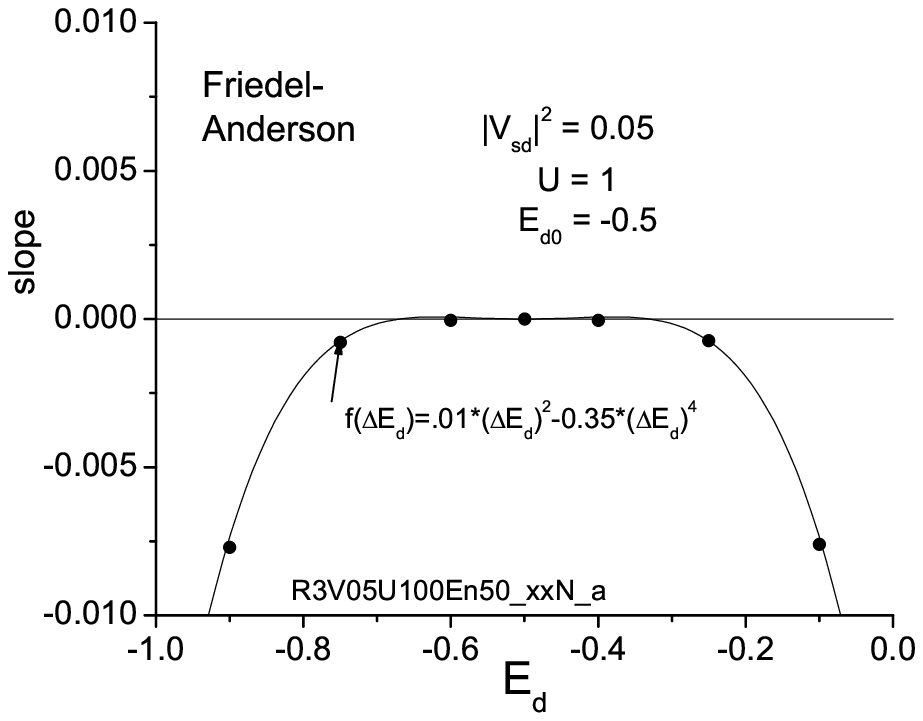}%
}%
\\
&
\begin{tabular}
[c]{l}%
Fig.6: The slope as shown in Fig.5 of $\ln\left(  F\right)  $ versus $N$ as a
function\\
of $E_{d}$ of the fidelity state for the asymmetric FA impurity.
\end{tabular}
\end{align*}

\subsection{Friedel impurity}

The FA impurity is defined by three parameters. Its fidelity is independent of
$N$ along the path $U=\lambda U_{0},E_{d}=-\frac{\lambda}{2}U_{0}$ and
(slightly) singular along other paths, for example along the path where only
$E_{d}$ is varied. This raises the question whether the singular behavior of
$\ln\left(  F\right)  $ is a consequence of the Coulomb interaction and the
resulting Kondo ground state or whether it is a trivial result of the single
particle potentials $V_{sd}$ and $E_{d}$. Therefore it is an obvious necessity
to check this question. Such a check is easy done by investigating the simple
spinless Friedel impurity which is defined by\ the two parameters $\left\vert
V_{sd}\right\vert ^{2}$ and $E_{d}.$

In the first Friedel investigation we choose for the nullstate the parameters
$\left\vert V_{sd}\right\vert ^{2}=0.05$ and $E_{d,0}=0$. Then a series of
fidelity series are performed with the same value for $\left\vert
V_{sd}\right\vert ^{2}$ and values for $E_{d}$ between $-0.7$ and $+0.7$. The
plots of $\ln\left(  F\right)  $ versus $N$ yield straight lines with a
relatively small slope. These slopes are plotted in Fig.7 as a function of
$E_{d}$. At $E_{d}=0$ the slope is, of course, zero because both states are
identical. As a whole one obtains a bell-shaped curve for the slopes. This
demonstrates that a simple change of $E_{d}$ yields a singular $\ln\left(
F\right)  $ for large $N$ without any electron-electron interaction.%

\begin{align*}
&
{\includegraphics[
height=3.2312in,
width=4.0747in
]%
{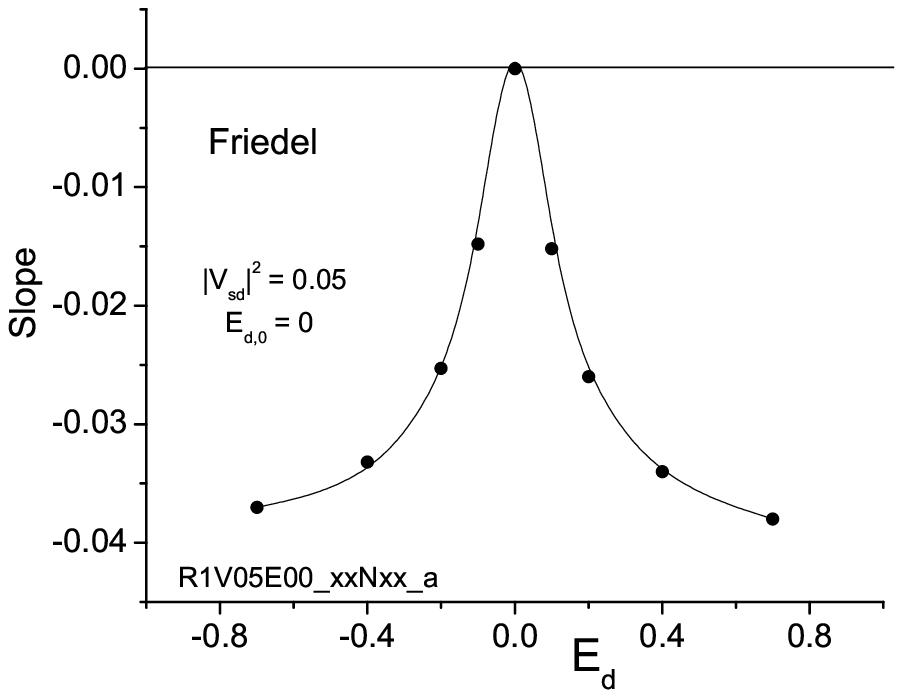}%
}%
\\
&
\begin{tabular}
[c]{l}%
Fig.7: The slope of $\ln\left(  F\right)  $ versus $N$ for the asymmetric
Friedel\\
impurity as a function of $E_{d}$. The nullstate is a Friedel impurity with\\
$E_{d}=0$ and the same $\left\vert V_{sd}\right\vert ^{2}=0.05$.
\end{tabular}
\ \ \ ^{{}}%
\end{align*}

In a second series of simulations the same nullstate is chosen with
$\left\vert V_{sd}\right\vert ^{2}=0.05$ and $E_{d,0}=0$. For the fidelity
states the s-d-hopping matrix $\left\vert V_{sd}\right\vert ^{2}$ is varied
between $5\times10^{-5}$ and $0.05$. For each value of $\left\vert
V_{sd}\right\vert ^{2}$ the fidelity approaches a constant value for a
sufficiently large number of Wilson states. There is no singular behavior of
$\ln\left(  F\right)  $ as a function of $N$. Of course, the constant value of
the fidelity depends on $\left\vert V_{sd}\right\vert ^{2}$. This dependence
of $F$ is shown in Fig.8 as a function of $\ln\left(  \left\vert
V_{sd}\right\vert ^{2}\right)  $.
\begin{align*}
&
{\includegraphics[
height=3.2594in,
width=4.0473in
]%
{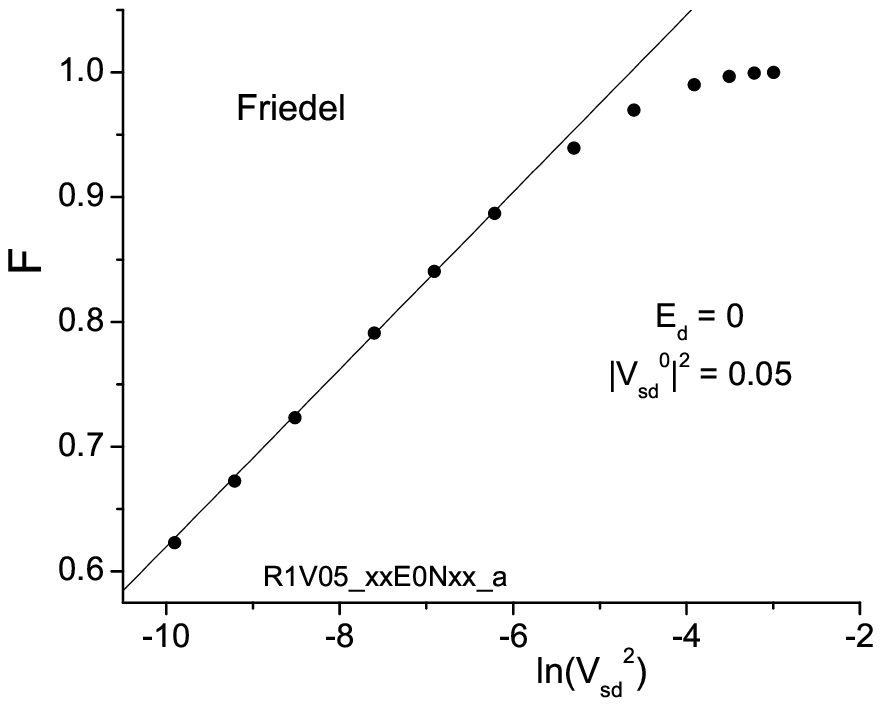}%
}%
\\
&
\begin{tabular}
[c]{l}%
Fig.8: Fidelity between a two Friedel states as a function of $\ln\left(
\left\vert V_{sd}\right\vert ^{2}\right)  $\\
in the fidelity state while the nullstate has $\left\vert V_{sd}\right\vert
^{2}=0.05$. Both states\\
have identical values of $E_{d}=0,$ $U=0$. The fidelity shows\\
essentially no dependence on the number of Wilson states $N$.
\end{tabular}
\end{align*}

\section{Discussion}

In our fidelity calculation we use the Wilson basis for the conduction band in
the (normalized) energy range $\left(  -1:+1\right)  $. The density of Wilson
states is very thin far away from the Fermi level and increases exponentially
close to the Fermi level. There is no question that the Wilson basis does not
describe well the density of states of a resonance far below or above the
Fermi level where the level separation is much larger than the resonance
width. Therefore it is a legitimate question whether this handicap of the
Wilson basis disqualifies it for fidelity calculations.

To clarify this question let us compare two different wave functions:
$\Psi_{1}$ describes the ground state of of a system with a resonance at
$\varepsilon_{d,1}=-0.8$ and $\Psi_{2}$ has a resonance at $\varepsilon
_{d,2}=-0.4$. Both resonances are sharp and have a width of $\Delta=0.05$. To
simplify the situation we assume that the s-d-matrix element vanishes for
$\left\vert \varepsilon-\varepsilon_{d,i}\right\vert >0.1$. In both systems
the conduction band is half filled (all states in the energy range $\left(
-1\leq\varepsilon\leq0\right)  $ are occupied). Intuitively one might assume
that the wave functions of $\Psi_{1}$ and $\Psi_{2}$ are quite different
because their resonances don't overlap. However, this is not the case. The
scalar product (fidelity) $\left\langle \Psi_{1}|\Psi_{2}\right\rangle $ is
essentially one. The reason is that in both wave functions the d-state and the
band states in the range $\left(  -1\leq\varepsilon\leq0\right)  $ are all
occupied so that their wave functions are given by
\[
\Psi_{1}\approx\Psi_{2}\approx\widehat{d}^{\dagger}%
{\displaystyle\prod\limits_{\varepsilon<0}}
\widehat{c}_{\varepsilon}^{\dagger}\Phi_{0}%
\]
where $\widehat{c}_{\varepsilon}^{\dagger}$ describes the band states with the
energy $\varepsilon$. The different density of states far below the Fermi
level (as well as far above) is not important for the fidelity. What counts in
the fidelity is the occupation of states close to the Fermi level. For this
reason the Wilson basis is particularly well suited for fidelity calculations
because it emphasizes the states close to the Fermi level where it counts and
it does not waste states far away from the Fermi level. In the appendix we
demonstrate that it is the (smallest) level separation at the Fermi level
which determines the fidelity. Halving the level separation by introducing one
additional state above and below the Fermi level has the same effect as
doubling the number of states (which also halves the level spacing at the
Fermi level).

An important question in this investigation is whether the fidelity identifies
and helps to understand interacting electron systems. Both the Kondo and the
FA impurities possess a singlet ground state. For sufficiently large Coulomb
repulsion between the spin-up and down impurity states the FA impurity shows a
behavior that is very similar to the Kondo impurity.

A comparison between Fig.1 for the Kondo impurity and Fig.3 for the FA
impurity shows that the fidelities of the two systems behave very differently.
For the following discussion it will be useful to calculate the
singlet-triplet excitation energy for the two systems. In table I the relaxed
singlet-triplet excitation energy $\Delta E_{st}$ is collected for the
parameters of the Kondo impurity investigated in \cite{B153}. The relaxed
singlet-triplet excitation energy $\Delta E_{st}$ is obtained by optimizing
the two bases $\left\{  \widehat{a}_{i}^{\dagger}\right\}  $ and $\left\{
\widehat{b}_{i}^{\dagger}\right\}  $ independently in the singlet state and
the triplet state.

For the development of the ground state it is important that the smallest
level separation $\delta E$ at the Fermi level (which is $\delta
E=2\ast2^{-N/2}$) is less than the excitation energy $\Delta E_{st}$.
Therefore we collect in table I also the critical number of Wilson states
$N_{st}\approx2\ast\left[  \left[  \log_{2}\left(  1/\Delta E_{st}\right)
\right]  +1\right]  $ that yields a level separation of about $\Delta E_{st}$.
In Fig.1 this critical value is marked with a small arrow. One recognizes that
for $N<N_{st}$ the fidelity is essentially constant and for $N>N_{st}$ the
logarithm of the fidelity changes linearly with $N$.
\[%
\begin{tabular}
[c]{|l|l|l|}\hline
$\mathbf{J}$ & $\mathbf{\Delta E}_{st}$ & $\mathbf{N}_{st}$\\\hline
$0.15$ & $9.1\times10^{-4}$ & $22$\\\hline
$0.12$ & $1.65\times10^{-4}$ & $26$\\\hline
$0.10$ & $2.53\times10^{-5}$ & $32$\\\hline
$0.09$ & $7.02\times10^{-6}$ & $36$\\\hline
$0.08$ & $1.52\times10^{-6}$ & $40$\\\hline
$0.07$ & $2.41\times10^{-7}$ & $44$\\\hline%
$<$%
$0.07$ &
$<$%
$10^{-7}$ & $>48$\\\hline
\end{tabular}
\ \ \ \ \ \ \
\]
Table I: The relaxed singlet-triplet excitation energy $\Delta E_{st}$ for the
Kondo impurity as a function of $J$. The third column gives the (closest)
number of Wilson states $N_{st}$ so that the smallest level separation is
roughly equal to the excitation energy $\Delta E_{st}$.
\[
\]

In table II the corresponding data $\Delta E_{st}$ and $N_{st}$ are collected
for different values of $\left\vert V_{sd}\right\vert ^{2}$ for the FA
impurity. In Fig.3 the critical values of $N_{st}$ are also marked on the
curves. However, now the behavior is almost reversed compared with the Kondo
impurity. For the FA impurity we observe essentially a linear decrease of
$\ln\left(  F\right)  $ with increasing $N$ for $N<N_{st}$ and a saturation of
$\ln\left(  F\right)  $ for $N>N_{st}$. In particular there is no singular
behavior of $\ln\left(  F\right)  $ for large $N$.%

\[%
\begin{tabular}
[c]{|l|l|l|}\hline
$\left\vert \mathbf{V}_{sd}\right\vert ^{2}$ & $\mathbf{\Delta E}_{st}$ &
$\mathbf{N}_{st}$\\\hline
$0.05$ & $8.33\times10^{-4}$ & $22$\\\hline
$0.04$ & $1.35\times10^{-4}$ & $26$\\\hline
$0.03$ & $3.23\times10^{-6}$ & $38$\\\hline
$0.025$ & $2.65\times10^{-7}$ & $44$\\\hline
\end{tabular}
\]

Table II: The relaxed singlet-triplet excitation energy $\Delta E_{st}$ as a
function of $\left\vert V_{sd}\right\vert ^{2}$. The third column gives the
(closest) number of Wilson states $N_{st}$ so that the smallest level
separation is roughly equal to the excitation energy $\Delta E_{st}$.%

\[
\]

This may be rather surprising since the symmetric FA impurity approaches the
Kondo impurity asymptotically for small $\left\vert V_{sd}\right\vert ^{2}/U$,
but this is not reflected by the fidelity behavior.

Recently Weichselbaum \textit{et al}. \cite{W50} calculated the fidelity of
the FA impurity using the numerical renormalization group (NRG) theory. They
obtained in general a logarithmic decrease of the fidelity. However, they used
very different fidelity paths. In one example they varied the energy of the
d-level and kept the other parameters constant. Therefore we performed a
similar calculation which is shown in Fig.5. We believe, however, that the
linear decrease of $\ln\left(  F\right)  $ with $N$ is not a many-body effect.
Therefore we have calculated the fidelity of the simple non-interacting
Friedel impurity. Fig.7 shows that one obtains a singular behavior of
$\ln\left(  F\right)  $ for large $N$. This is not surprising since it was
derived earlier by Anderson and is known as the Anderson orthogonality
catastrophe. We believe that the singular behavior as observed by Weichselbaum
\textit{et al}. is due to the change of the potential scattering in the
underlying Friedel resonance. Weichselbaum \textit{et al}. use rather small
parameters of $U,E_{d}$ and $\left\vert V_{sd}\right\vert ^{2}$ such as
$U=0.12$ and $\Gamma_{\mu}=\pi\left\vert V_{sd}^{\mu}\right\vert ^{2}\rho
_{\mu}$ $=0.01$ and several hybridization processes $\mu$ ($\rho_{\mu}$ is the
density in the band $\mu$). We did not extend our software to several
hybridization processes since we concluded that our two examples of the Kondo
and the FA impurity already illuminate the physics.

We suggest the following mechanisms for the different behavior of the fidelity
$\ln\left(  F\right)  $ as a function of $N$. In the Kondo impurity we compare
the Kondo solution with the $J=0$ state. The latter is a homogeneous electron
gas with the same density at the impurity as anywhere else. For small $J$ the
magnetic d-electron causes only a relatively small change for small $N$
\ since the Kondo ground state has not yet developed. When $N$ becomes larger
than $N_{st}$ the Kondo ground state has formed and causes a phase shift of
$\pi/2$ for all electrons with smaller energy. This phase shift is the reason
why the scalar product with the free electron case ($J=0$) goes to zero, i.e.
$\ln\left(  F\right)  $ diverges. It is analogous to the Anderson
orthogonality catastrophe.

The difference between the Kondo and the symmetric FA impurity is that we
don't compare the latter with the free electron case but with a state that has
the same s-d-potential $\left\vert V_{sd}\right\vert ^{2}$. If one chooses for
the nullstate the symmetric Friedel impurity then all electrons within the
resonance width already have a phase shift of $\pi/2$ in the nullstate. On the
other hand in the singlet ground state of the FA impurity all electrons with
energy smaller the $\Delta E_{st}$ also have a phase shift of $\pi/2$. Since
this is the same phase shift as in the nullstate it does not reduce the scalar
product of the fidelity between the nullstate and the singlet ground state
with increasing $N$. The fidelity becomes asymptotically constant.

Finally it is tempting to compare the ground-state wave function of a Kondo
impurity with that of a FA impurity. From tables I and II one finds that the
Kondo impurity with $J=0.12$ and the FA impurity with $\left\vert
V_{sd}\right\vert ^{2}=0.04$, $U=1$ and $E_{d}=-0.5$ have roughly the same
singlet-triplet excitation energy ($1.65\times10^{-4}$ versus $1.35\times
10^{-4}$). Therefore we calculate the scalar product which yields the
similarity between the wave functions for different $N$. This similarity
(which is defined in the literature as fidelity) is plotted in Fig.9. It shows
that $F$ is close to 1.0 and approaches a constant value of $0.95$ for large
$N$. This confirms the similarity between the Kondo and the FA impurity (for
large $U/\left\vert V_{sd}\right\vert ^{2}$), and the phase shift in both
systems close to the Fermi level is essentially the same.
\begin{align*}
&
{\includegraphics[
height=3.1474in,
width=4.0473in
]%
{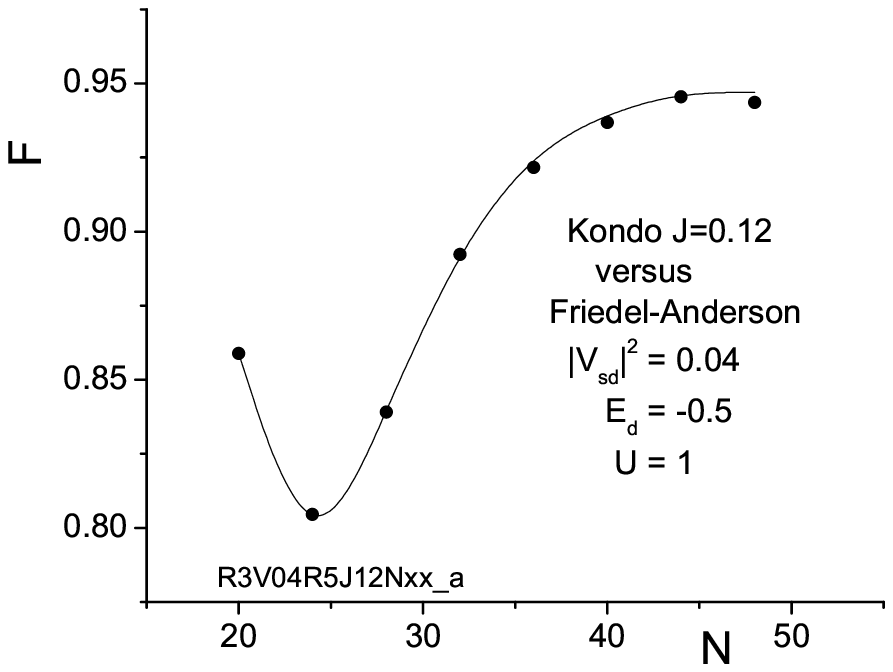}%
}%
\\
&
\begin{tabular}
[c]{l}%
Fig.9: The similarity (fidelity) between the ground state of\\
a Kondo and a FA impurity as a function of the number of\\
Wilson states. The parameters of the impurities are shown\\
in the figure.
\end{tabular}
\end{align*}

The minimum of the curve is at about $N=24$. This corresponds to a level
separation at the Fermi energy of $2\ast2^{-N/2}\approx\allowbreak
1.2\times10^{-4}$. This is of the order of the singlet-triplet excitation
energy of the two systems.

\newpage

\section{Conclusion}

In this paper the ground states of the Kondo impurity and the Friedel-Anderson
impurity are calculated for many parameters and seven different numbers $N$ of
Wilson states using the FAIR theory. The effective number of band electrons is
$N_{eff}\approx2\ast2^{N/2}$. For each number of Wilson states the resulting
ground states\ (which we denote as fidelity states) are compared with the
corresponding ground states for zero interaction, the so-called nullstates.
The fidelity is obtained by forming the scalar product between the fidelity
state and the nullstate for each $N$. Then the logarithm of the fidelity
$\ln\left(  F\right)  $ is plotted versus $N\approx2\log_{2}\left(
N_{eff}/2\right)  $.

The fidelity shows very different behavior for the Kondo and the
Friedel-Anderson impurities. In the symmetric FA impurity it saturates at
large values of $N$ while for the Kondo impurity the logarithm $\ln\left(
F\right)  $ diverges. This result demonstrates that the behavior of the
fidelity depends as much on the choice of the simple nullstate as on the
interacting fidelity state.

For the symmetric FA impurity we choose a nullstate with $U=0$ and $E_{d}=0$
(to maintain the symmetry) but leave $\left\vert V_{sd}\right\vert ^{2}$
constant. Here the s-electrons close to the Fermi energy already have a phase
shift of $\pi/2$ in the nullstate. The interacting ground state introduces a
phase shift of $\pi/2$ as well in a narrow energy range about the Fermi
energy. Therefore if one increases the number of states closer and closer to
the Fermi level the phase shift in the nullstate and the fidelity state are
the same and the scalar product does not change. On the other hand, for the
Kondo impurity we set $J=0$ and obtain a nullstate whose conduction band is
the free electron band that has no phase shift, and the fidelity decreases
with increasing $N$. It is not sufficient to turn off the interaction in the
nullstate. One also has to know or investigate the phase shift of its
s-electrons close to the Fermi level.

In this respect the fidelity calculations yield comparative information about
the s-electrons at the Fermi level. In addition a change in the slope of
$\ln\left(  F\right)  $ versus $N$ indicates at which energy the inner
structure changes, either of the nullstate or the fidelity state.

If we compare two multi-electron states then the behavior of the fidelity does
not tell us whether none, one or both are interacting electron systems. The
fidelity does not correlate with the many-body physics of the problems.

Finally, we observed that the fidelity of ground states of the Kondo and the
FA Hamiltonians with similar Kondo temperatures does not show an Anderson
orthogonality catastrophe, but on the contrary is relatively close to one and
becomes constant with an increasing number of Wilson states $N.$

\newpage

\section{Appendix}

\appendix{}

\section{Wilson's states}

Wilson considered an s-band with a constant density of states and the Fermi
energy in the center of the band. By measuring the energy from the Fermi level
and dividing all energies by the Fermi energy Wilson obtained a band ranging
from $-1$ to $+1$. To treat the electrons close to the Fermi level at
$\zeta=0$ as accurately as possible he divided the energy interval $\left(
-1:0\right)  $ geometrically at energies of $\zeta_{\nu}=\Lambda^{-\nu}$. In
most cases the value $\Lambda=2$ is used yielding $-1/2,-1/4,-1/8,.$. i.e.
$\zeta_{\nu}=-1/2^{\nu}$. This yields energy cells $\mathfrak{C}_{\nu}$ with
the range $\left\{  -1/2^{\nu}:-1/2^{\nu+1}\right\}  ,$ width $\Delta_{\nu}$
$=\zeta_{\nu+1}-\zeta_{\nu}$ $=1/2^{\nu+1}$ and average energy $\varepsilon
_{\nu}=\left(  \zeta_{\nu}+\zeta_{\nu-1}\right)  /2$.

Wilson rearranged the quasi-continuous original electron states $\varphi
_{k}\left(  x\right)  $ in such a way that only one state within each cell
$\mathfrak{C}_{\nu}\ $had a finite interaction with the impurity. Assuming
that the interaction of the original electron states $\varphi_{k}\left(
x\right)  $ with the impurity is independent of $k$, this interacting state in
$\mathfrak{C}_{\nu}$ had the form%
\[
\psi_{\nu}\left(  x\right)  =%
{\textstyle\sum_{\mathfrak{C}_{\nu}}}
\varphi_{k}\left(  x\right)  /\sqrt{Z_{\nu}}%
\]
where $Z_{\nu}$ is the total number of states $\varphi_{k}\left(  x\right)  $
in the cell $\mathfrak{C}_{\nu}$ ($Z_{\nu}=Z\left(  \zeta_{\nu+1}-\zeta_{\nu
}\right)  /2,$ $Z$ is the total number of states in the band). There are
$\left(  Z_{\nu}-1\right)  $ additional linear combinations of the states
$\varphi_{k}$ in the cell $\mathfrak{C}_{\nu}$ but they have zero interaction
with the impurity and were ignored by Wilson as they are within this paper.

The interaction strength of the original basis states $\varphi_{k}\left(
x\right)  $ with the d-impurity is assumed to be a constant, $v_{sd}$. Then
the interaction between the d-state and the Wilson states $\psi_{\nu}\left(
x\right)  $ is given by $V_{sd}\left(  \nu\right)  =V_{sd}^{0}\sqrt{\left(
\zeta_{\nu+1}-\zeta_{\nu}\right)  /2}$ where $\left\vert V_{sd}^{0}\right\vert
^{2}=$ $%
{\textstyle\sum_{k}}
\left\vert v_{sd}\right\vert ^{2}=$ $%
{\textstyle\sum_{\nu}}
\left\vert V_{sd}\left(  \nu\right)  \right\vert ^{2}.$

\subsection{FAIR theory}

Let us first consider the Friedel impurity without spin. Its Hamiltonian is
\begin{equation}
\widehat{H}_{F}=%
{\displaystyle\sum_{\nu=1}^{N}}
\varepsilon_{\nu}\widehat{c}_{\nu}^{\dagger}\widehat{c}_{\nu}+E_{d}\widehat
{d}^{\dagger}\widehat{d}_{\sigma}+%
{\displaystyle\sum_{\sigma}}
V_{\nu}^{sd}\left(  \widehat{c}_{\nu}^{\dagger}\widehat{d}_{\sigma}%
+\widehat{d}^{\dagger}\widehat{c}_{\nu}\right)  \label{H_F}%
\end{equation}
We call this Hamiltonian sub-diagonal because it is diagonal in the states
$\widehat{c}_{\nu}^{\dagger}$ but not between $\widehat{c}_{\nu}^{\dagger}%
$\ and $\widehat{d}^{\dagger}$. (We use here the creation operators to denote
the corresponding states $\widehat{c}_{\nu}^{\dagger}\Phi_{0}$ or $\widehat
{d}^{\dagger}\Phi_{0}$, where $\Phi_{0}$ is the vacuum).

By diagonalization one finds the exact eigenstates
\begin{equation}
\widehat{b}_{j}^{\dagger}=%
{\displaystyle\sum_{\nu=1}^{N+1}}
\beta_{j}^{\nu}\widehat{c}_{\nu}^{\dagger}+\beta_{j}\widehat{d}^{\dagger}
\label{b_j}%
\end{equation}
and a diagonal Hamiltonian. The ground state with $n$ electrons is given by
\begin{equation}
\Psi_{F}=%
{\displaystyle\prod\limits_{j=1}^{n}}
\widehat{b}_{j}^{\dagger}\Phi_{0} \label{Psi0_F}%
\end{equation}
where $\Phi_{0}$ is the vacuum state.

Of course, one can reverse the process and starting from the diagonal
Hamiltonian $\widehat{H}=%
{\displaystyle\sum_{j}}
E_{j}^{b}\widehat{b}_{j}^{\dagger}\widehat{b}_{j}$ extract the resonance state
$\widehat{d}^{\dagger}$ and build an arbitrary orthonormal basis out of the
$\widehat{b}_{j}^{\dagger}$ which is orthogonal to $\widehat{d}^{\dagger}$.
The Hamiltonian will not be diagonal in this basis. So in the final step one
sub-diagonalizes the Hamiltonian excluding the state $\widehat{d}^{\dagger}$
in the process.

This reverse process can also be applied to the s-electron part of
$\widehat{H}_{F}$. One can build an arbitrary state $\widehat{a}_{0}^{\dagger
}=%
{\displaystyle\sum_{\nu}}
\alpha_{0}^{\nu}\widehat{c}_{\nu}^{\dagger}$. In the next step one builds a
new orthonormal conduction band basis $\left\{  \widehat{a}_{i}^{\dagger
}\right\}  $ with $\left(  N-1\right)  $ states which are also orthogonal to
$\widehat{a}_{0}^{\dagger}$. Again the Hamiltonian $\widehat{H}^{0}$ will not
be diagonal and in the final step one sub-diagonalizes the Hamiltonian
excluding the state $\widehat{a}_{0}^{\dagger}$ in the process. Now
$\widehat{a}_{0}^{\dagger}$ is an artificial Friedel resonance, i.e. the FAIR
state. The state $\widehat{a}_{0}^{\dagger}$ determines the composition of the
whole basis $\left\{  \widehat{a}_{i}^{\dagger}\right\}  $.

This FAIR concept is rather flexible because $\widehat{a}_{0}^{\dagger}$ can
be any combination of the s-states $\widehat{c}_{\nu}^{\dagger}$. It turns out
that there is one special state $\widehat{a}_{0}^{\dagger}$ with which one can
construct the exact ground state of the Friedel resonance. With this special
FAIR state the Friedel ground state takes the form%
\[
\Psi_{F}=\left(  A\widehat{a}_{0}^{\dagger}+B\widehat{d}^{\dagger}\right)
{\displaystyle\prod\limits_{i=1}^{n-1}}
\widehat{a}_{i}^{\dagger}\Phi_{0}%
\]

This ground state of the Friedel resonance has the great advantage that
$\widehat{d}^{\dagger}$ is only hybridized with one single s-electron
$\widehat{a}_{0}^{\dagger}$. The FAIR state $\widehat{a}_{0}^{\dagger}$ is in
a way representing all other s-electrons. $\left(  A\widehat{a}_{0}^{\dagger
}+B\widehat{d}^{\dagger}\right)  $ forms a composed state which shifts the
energy of all the other electron states (introducing a phase shift). It is the
building block for the compact ground state of the FA and the Kondo impurity.

\section{Relation between Wilson state number $N$ and effective number of
electrons $N_{eff}$}

The Wilson states are defined by the ratio $\Lambda$. However, for some
physical properties this sub-division of the energy band is too coarse. We
observed an error in the amplitude of the Friedel oscillation of about 10\%
for $\Lambda=2$ which became of the order of 1\% for when the intervals where
sub-divided twice (corresponding to $\Lambda=\sqrt[4]{2}$ \cite{B178}.
Therefore we checked whether the coarse sub-division of the band caused any
error for the fidelity calculation. For this purpose we calculated the
fidelity between two Friedel resonances with two different d-energies,
$E_{d1}=0$ and $E_{d2}=-1$. In both cases the s-d-coupling is $\left\vert
V_{sd}\right\vert ^{2}=0.05$. In Fig.10 the squares give the plot of
$\ln\left(  F\right)  $ versus the number of Wilson states for $\Lambda=2$,
which is equivalent to all the plots in this paper. Then we subdivided each
cell into two equal subcells (full circles) and again each subcell into two
new subcells (full triangles). The number of states increased each time by a
factor two but we plotted the newly calculated $\ln\left(  F\right)  $ as a
function of the original number $N$ of Wilson states. First we observe that
the resulting straight lines are perfectly parallel. Secondly the two
sub-divisions into equal subcells reduced the smallest energy $\delta E$ at
the Fermi level by a factor of $4=2^{2}$. This means that after two
subdivisions the smallest $\delta E$ for $N$ Wilson states is equal to the
original $\delta E$ for $\left(  N+4\right)  $ Wilson states. For example, the
plot shows that the square at $N=36$ has the same value as the triangle at
$N=32$. If one would plot $\ln\left(  F\right)  $ versus $2\log_{2}\left(
1/\delta E\right)  $ all points would fall on one straight line (the one with
the squares), although the number of states used in the calculation are varied
by a factor four. This demonstrates that the fidelity depends essentially on
the smallest energy $\delta E$ at the Fermi energy and not on the total number
of states.
\begin{align*}
&
{\includegraphics[
height=3.2312in,
width=3.8082in
]%
{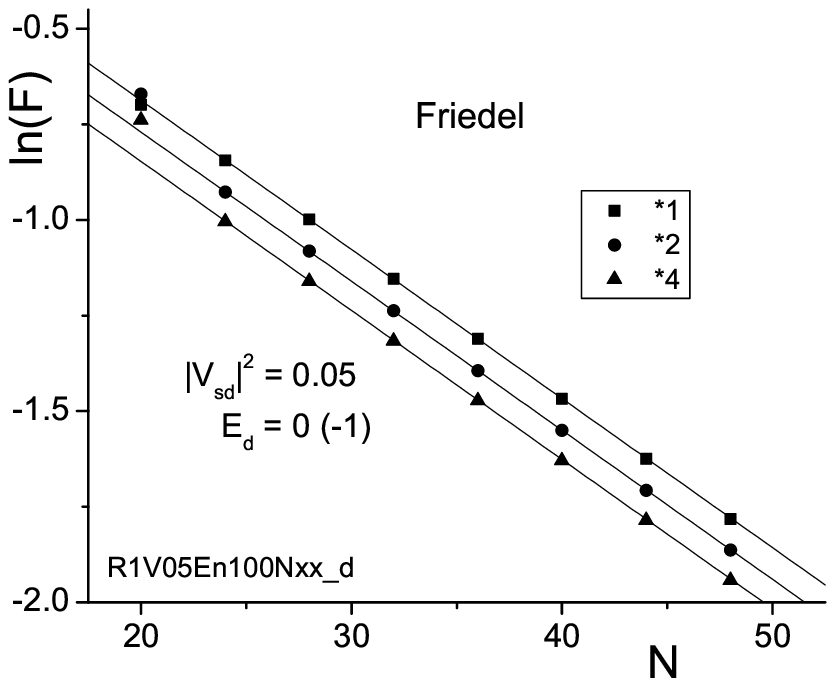}%
}%
\\
&
\begin{tabular}
[c]{l}%
Fig.10: The logarithm of the fidelity between two Friedel impurities\\
with different d-level energies. The squares are for a regular Wilson\\
spectrum. For the circles each Wilson energy cell is divided into two\\
cells increasing the number of states to $2N$. For the triangles the\\
original energy cells are divided into four cells yielding $4N$ states.\\
Therefore the new states are essentially a factor 2 and 4 closer, and\\
the smallest energy separation is smaller\ by a factor 2 and 4. The\\
fidelity is plotted in all cases versus the original number of Wilson\\
states. The straight lines are perfectly parallel. In addition a triangle\\
at $N=36$ has the same smallest energy as a square at $N=40$, and\\
indeed they have the same fidelity.
\end{tabular}
\end{align*}

\end{document}